\documentclass[a4paper]{article}
\usepackage{indentfirst}
\usepackage{hyperref}
\usepackage{multirow}
\usepackage{latexsym,bm,amsmath,amssymb}
\usepackage{graphicx}
\usepackage{epsfig}
\usepackage{subfigure}
\usepackage{cite}
\usepackage{color}
\usepackage[top=1in, bottom=1in, left=1.25in, right=1.25in]{geometry}

\usepackage{tikz}
\usetikzlibrary{shapes.geometric, arrows}
\usetikzlibrary{trees}
\usetikzlibrary{decorations.pathmorphing}
\usetikzlibrary{decorations.markings}
\tikzset{
        photon/.style={decorate, decoration={snake}, draw=red},
        nucleon/.style={draw=black, postaction={decorate},
           decoration={markings,mark=at position .55 with{\arrow[draw=black]{>}}}},
        pion/.style={draw=blue, postaction={decorate},
        decoration={markings,mark=at position .55 with{\arrow[draw=blue]{}}}},
        sigma/.style={draw=black, postaction={decorate},
        decoration={markings,mark=at position .55 with{\arrow[draw=black]{}}}},
        link/.style    = { draw=black, double = white, line width = 1.8pt, double distance = 0.8pt , postaction={decorate},decoration={markings,mark=at position .55 with{\arrow[draw=black]{>}}}},
    }

\def\bea{\arraycolsep .1em \begin{eqnarray}}
\def\eea{\end{eqnarray}}

\def\s0#1#2{\mbox{\small{$ \frac{#1}{#2} $}}}
\def\0#1#2{\frac{#1}{#2}}

\begin{document}

\setcounter{topnumber}{10}
\setcounter{totalnumber}{50}
\title{How does the $S_{11}$ $N^*(890)$ state emerge from a naive $K$ matrix fit?}
\maketitle

\begin{center}
{\sc Yao~Ma,$^{\dagger\,}$\footnote{e-mail address:mymz825@163.com}\,\,\, Wen-Qi~Niu,$^{\dagger\,}$\footnote{e-mail address:1701110076@pku.edu.cn}\,\,\, Yu-Fei~Wang,$^\ddagger$\footnote{e-mail address:yfwang@hiskp.uni-bonn.de}\,\,\,
Han-Qing~Zheng$^{\dagger\,,\star\,,}$\footnote{e-mail address:zhenghq@pku.edu.cn}}
\\
\vspace{0.5cm}

\noindent{\small{$^\dagger$ \it  Department of Physics and State Key
Laboratory of Nuclear Physics and Technology,
 Peking University, Beijing 100871, P.~R.~China}}\\
\noindent{\small{$^\ddagger$ \it  Helmholtz Institut f\"{u}r Strahlen- und Kernphysik, Bonn University, Bonn 53115, Germany}}\\
\noindent{\small{$^\star$ \it   Collaborative Innovation Center of
Quantum Matter, Beijing, Peoples Republic of China}}
\end{center}
\begin{abstract}
We give a pedagogical analysis on $K$ matrix models describing the $\pi N$ scattering amplitude, in $S_{11}$ channel at
low energies. We show how the correct use of analyticity in the $s$ channel and crossing symmetry in $t$ and $u$ channels leads to a much improved analytic behavior in the negative $s$ region, in agreement with the prediction from chiral perturbation amplitudes in its validity region. The analysis leads again to the conclusion that a genuine $N^*(890)$ resonance exists.
\end{abstract}

\vspace{1cm}

In a series of recent publications~\cite{Wang:2017agd,Wang:2018gul,Wang:2018nwi}, it is suggested that there exists a subthreshold resonance in the $S_{11}$ channel of $\pi N$ scattering, with $M-i\Gamma/2= 895(81)-i164(23)$MeV. The result is completely novel and is obtained based on an approach~(\cite{Zheng:2003rw,Zhou:2004ms,Zhou:2006wm}) with full respects to fundamental principals of $S$-matrix theory such as  unitarity, analyticity and consistent with crossing symmetry~(\cite{Guo:2007ff, Guo:2007hm}). The approach is much superior to conventional unitarization approximations, such as $K$-matrix method or variations of Pad\'e approximation (for the discussion, see for example Refs.~\cite{Qin:2002hk,Guo:2006br}).

Specifically, in the regime of Peking University (PKU) representation, Refs.~\cite{Wang:2017agd,Wang:2018gul,Wang:2018nwi} input the left-hand cut from chiral perturbation theory results, while the inelasticity and known poles are fixed from experimental data. Due to the negative definite contributions from left-hand cut to the phase shift, and positive definite contributions from known resonances and inelasticity, we claim that a sub-threshold resonance must exist, irrespective to the details of calculations, in $S_{11}$ channel. Especially, Ref.~\cite{Wang:2018nwi} gives a comprehensive analysis on not only the $N^*(890)$ pole but also the physics of all other $S$- and $P$-wave channels.
	
Nevertheless, those works have still received some queries: firstly, why do many previous works do not find significant effects from $N^*(890)$ resonance? Secondly, Refs.~\cite{Wang:2017agd,Wang:2018gul,Wang:2018nwi} inputs the left-hand cut only from chiral perturbation theory, which may not work well in the region far away from threshold.

The aim of this paper is to make a further analysis in the conventional $K$-matrix approach, to examine and to understand why the $N^*(890)$ was missed in previous researches, and how it can appear even in a simple $K$-matrix analysis, if one be respectful to the analyticity property of scattering amplitude offered by standard quantum field theory. We hope the analysis made in this paper be helpful to convince the physics community to accept
the existence of the $N^*(890)$ resonance.

To begin with, let us start from the standard coupled channel $K$-matrix formula:
\begin{equation}\label{case1-kmatrix}
T^{-1}=K^{-1}-i\rho \ ,
\end{equation}
where $\rho$ is a diagonal matrix with its elements being the kinematical factors in each channel, i.e,
$\rho=diag(\rho_1,\rho_2)$. To be specific, the subscripts 1,2 means $\pi N$  and  $\eta N$ channels, respectively.
Hence
\begin{equation}
\begin{aligned}
\rho_1=&\sqrt{[s-(m_N+m_\pi)^2][s-(m_N-m_\pi)^2]}/s\ , \\
 \rho_2=&\sqrt{[s-(m_N+m_\eta)^2][s-(m_N-m_\eta)^2]}/s\ .
\end{aligned}
\end{equation}

Basically the only requirement on $K$ is that it is a real symmetric matrix in the physical region in order to fulfil unitarity. Hence we start from a simple form (herewith we call it \textbf {Fit I}):
\begin{equation}\label{fit1}
K_{ij}=\sum_{\alpha=1,2}\frac{c^{\alpha}_{ij}}{s-M^2_{\alpha}}+P^{(2)}_{ij}(s)\ .
\end{equation}
In the above parametrization we put two bare poles with {masses} $M_1$ and $M_2$ because we knew that there are two well established resonance $N^*(1535)$ and $N^*(1650)$ to be adjusted, in  $S_{11}$ channel; and $P^{(2)}_{ij}(s)$ are 2$nd$ order real polynomials to simulate the background, and are symmetric in subscripts $i$ and $j$ (so are the coefficients $c^\alpha_{ij}$).

\begin{figure}[htbp]
\center
\subfigure[]{
\scalebox{1.2}[1.2]{\includegraphics[width=0.35\textwidth]{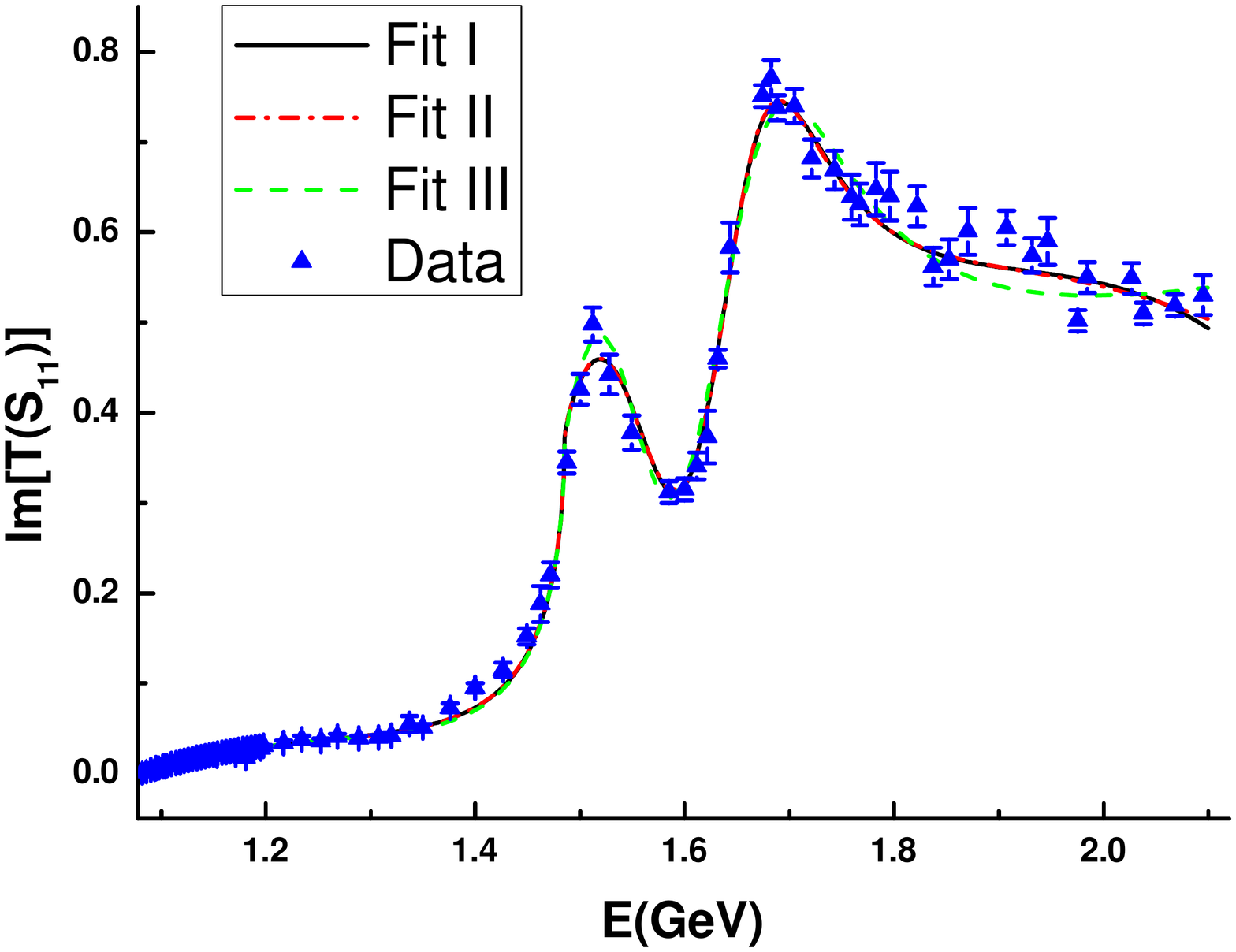}}
} \ \ \ \
\subfigure[]{
\scalebox{1.2}[1.2]{\includegraphics[width=0.35\textwidth]{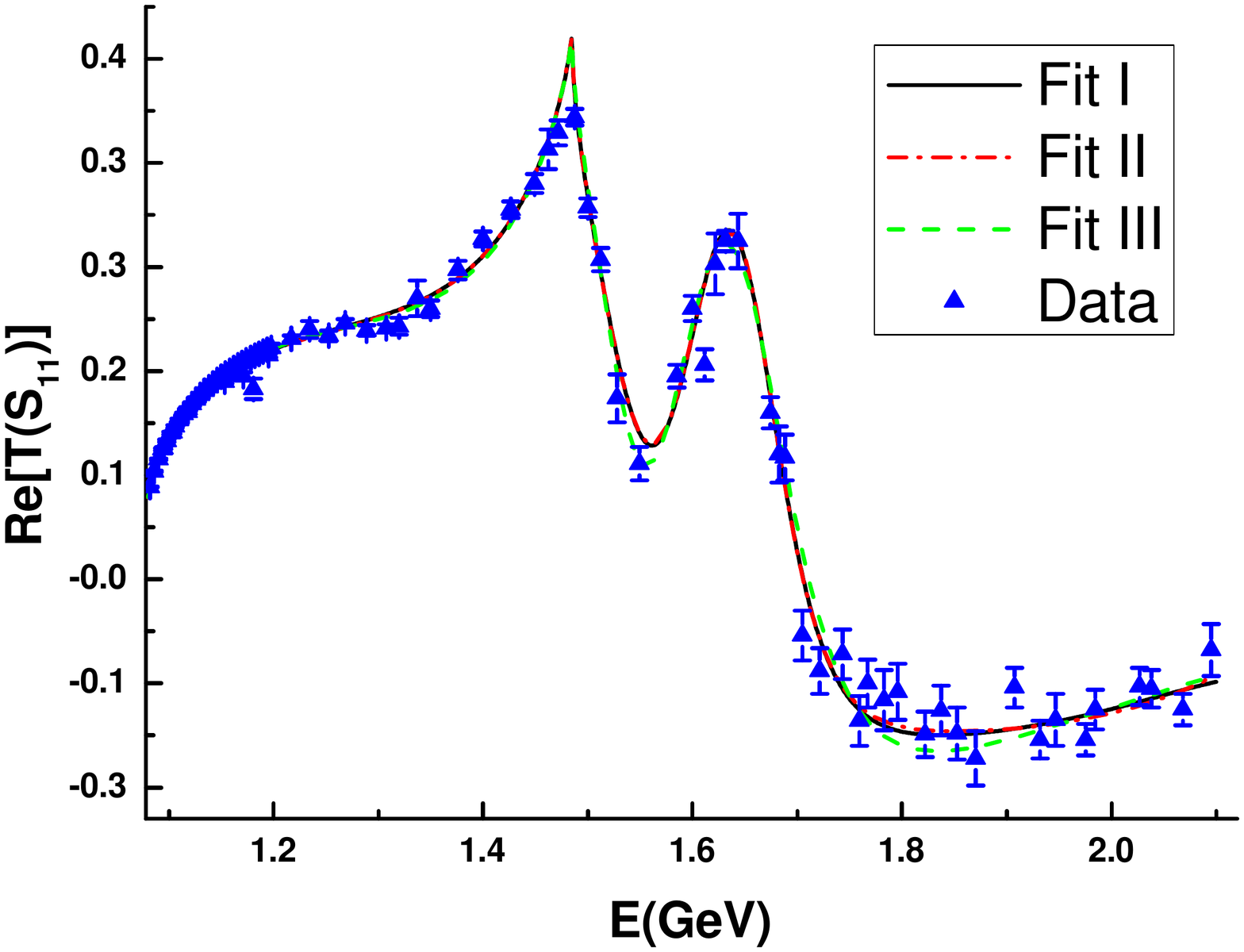}}
}
\caption{ Fit of GWU data upto $\sqrt{s}=2.1$GeV,  a) $\mathrm{Im} T$,
b) $\mathrm{Re} T$. Fit I: Solid line (black); Fit II: dot - dashed line (red); Fit III: dashed line (green). }\label{figcase1}
\end{figure}

Using Eq.~(\ref{fit1}) we fit $S_{11}$ partial wave data of $\pi N$ scatterings from threshold upto $\sqrt{s}=2.1$GeV provided by Ref.~\cite{Arndt:1985vj}; and to constrain the near-threshold behavior, we also fit the data of low-energy region in Ref.~\cite{Hoferichter:2015hva} simultaneously. The fit is plotted in Fig.~\ref{figcase1}. In practice, the $K$-matrix fit is notoriously known as producing spurious physical sheet resonance (SPSR) poles, but when they (and normal poles) are of less significance, i.e. contributing to the phase shift tinily, people do not care about them. The poles found from Fit I are listed in Table~\ref{tabcase1}. Besides, the scattering length is $a\simeq 162.43\times 10^{-3}m_\pi^{-1}$, which is compatible with, e.g. the result of Roy-Steiner analyses in Ref.~\cite{Hoferichter:2015hva}: $a\simeq 169.9(19.4)\times 10^{-3}m_\pi^{-1}$.

\begin{figure}[h]
	\centering
	\includegraphics[width=0.65\textwidth]{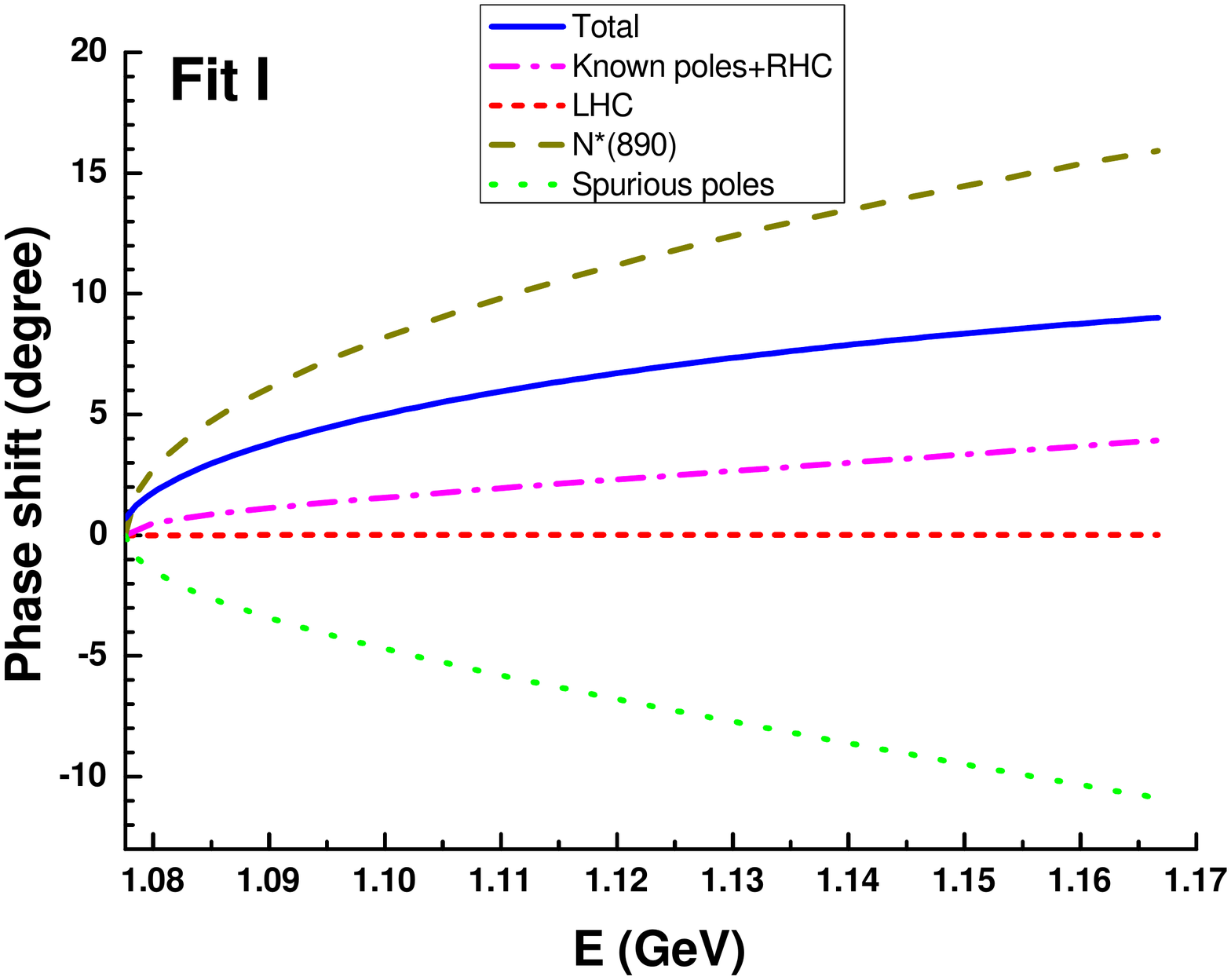}
	\caption{Fit I: Different contributions to the phase shift near $\pi N$ threshold.}
	\label{phase-1}
\end{figure}

Table~\ref{tabcase1} may deserve a few words of explanation: experienced {readers} will quickly recognize that poles $1.68- i 0.07$(II) and $1.68- i 0.07$(III) are of $N^*(1650)$, whereas $1.529- i 0.016$(IV) and $0.92$(III) are of $N^*(1535)$ since, as is well known, the former mainly couples to $\pi N$ while the latter mainly couples to $\eta N$.
Though their pole locations may be rather poorly determined,  the main interests here is however to investigate the $N^*(890)$ pole. Hence we actually do not pay much attention to these well established resonances. Similar to Refs.~\cite{Wang:2017agd,Wang:2018gul,Wang:2018nwi}, in Fig.~\ref{phase-1} phase shifts contributed from different sources are plotted using PKU representation; their sum equals to the fit curve or experimental data, as it should.
\vspace{0cm}
\begin{table}[h]
    \caption{Fit I: Poles obtained using Eq.~(\ref{fit1})}
    \centering
    \begin{tabular}{p{2cm}p{8cm}}
        \hline
         Sheets & pole positions on $\sqrt{s}$ plane in unit of GeV \\
        \hline
		$\mathrm{I}$              & $1.36 - i0.65$ \\
	  \hline	
$\mathrm{II}$                 & $1.68- i 0.07$; $0.73- i 0.19$   \\
	  \hline	$\mathrm{III}$              & $1.68- i 0.07$; $0.92$; 
\\
  \hline		$\mathrm{IV}$              & $1.529- i 0.016$ 
  \\
		\hline
    \end{tabular}
    \label{tabcase1}
\end{table}

We summarize major outputs from Fit I:
  \begin{itemize}
    \item The known poles and right hand inelastic cut which begins at the $\eta N$ threshold cannot fit the phase shift data. That indicates missing contributions from other sources have to exist.
    \item The left hand cut contribution is nearly zero, {which} is of course not correct as will be discussed at some lengths below.
    \item Below $\pi N$ threshold there exists a resonance pole providing a large positive phase shift. Meanwhile a spurious pole on first sheet provides a negative phase shift.
  \end{itemize}

\begin{figure}[h]
	\centering
	\includegraphics[width=0.65\textwidth]{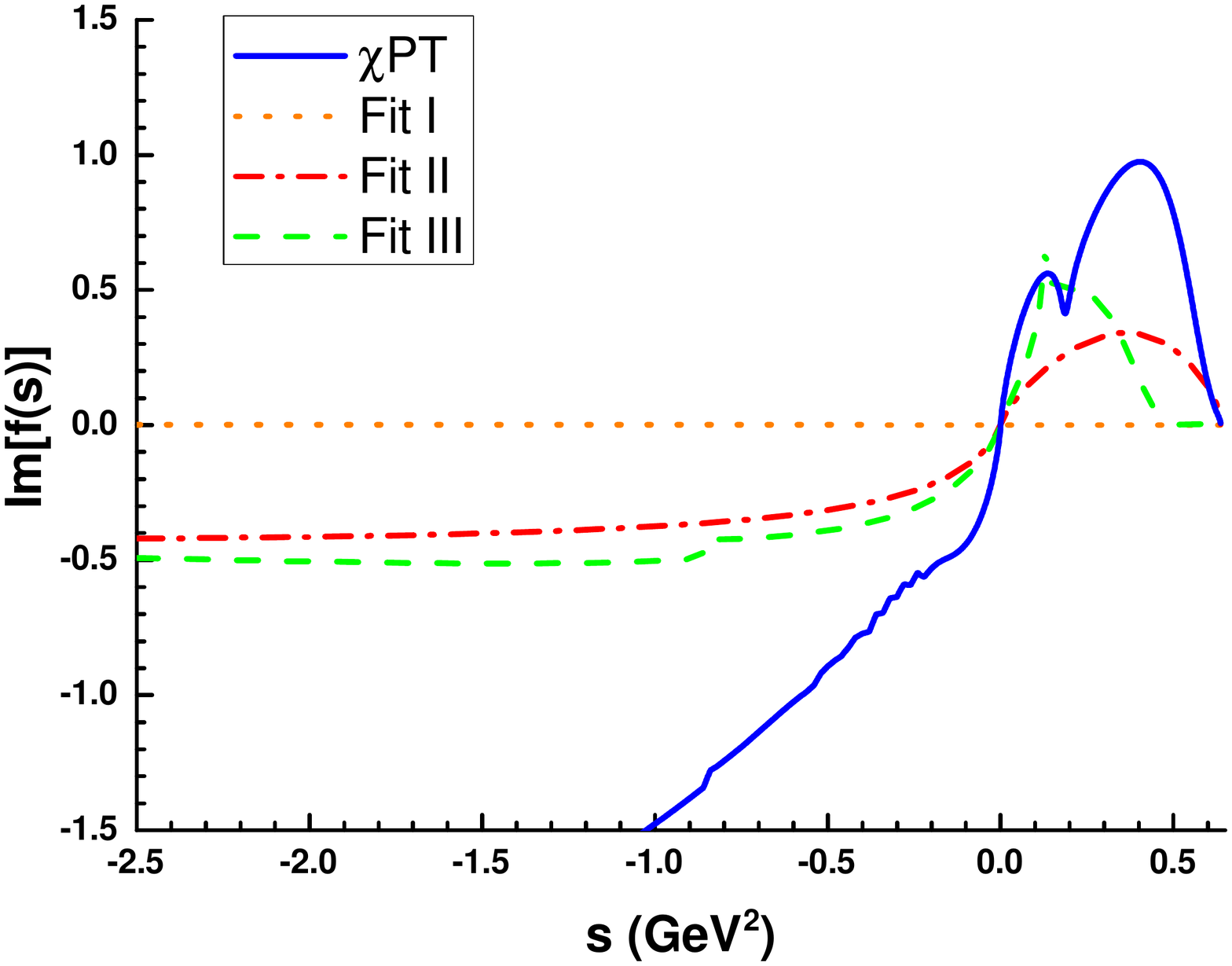}
	\caption{Different estimations on the {spectral function} of left hand integral: The solid (blue) line depicts the $\chi PT$ result at $O(p^3)$~\cite{Wang:2018nwi};
the dotted straight line (orange) is from Fit I; dot-dashed (green) line is from Fit II; dashed (green) line is generated by Fit III. Small fluctuations in these curves are either from nearby (spurious) singularities or from numerical instabilities and should be ignored.}
	\label{LHC-1}
\end{figure}

One great advantage of the PKU representation is that different contributions to the phase shift are separable and additive, hence one can calculate different contributions to the phase shift, especially that from the background term {in $\pi N$ channel, $f(s)\equiv\frac{\ln S_{11}(s')}{2i\rho_1(s')}$, with $S_{11}\equiv 1+2i\rho_1 T_{11}$:\footnote{For more discussions on related topics, one is referred to Refs.~\cite{Wang:2017agd,Wang:2018gul,Wang:2018nwi}. }
\begin{equation}\label{f}
	f(s) = \frac{s}{\pi}\int_{L}ds'\frac{\mathrm{Im}(\frac{\ln S_{11}(s')}{2i\rho_1(s')})}{s'(s'-s)}\ .
\end{equation}}
The {spectral function} in above integral, {$\mathrm{Im} f(s')=\mathrm{Im}(\frac{\ln S_{11}(s')}{2i\rho_1(s')})$} read off from Fit I is plotted in Fig.~\ref{LHC-1}. 
In the non-relativistic limit, function $\rho_1(s)f(s)$ reduces to $-kR$~\cite{Hu:1948zz} (see the discussion in {Ref.~\cite{Wang:2018nwi}})
with $k$ the center of mass momentum, and {$R=\frac{-2f(s_{R})}{(m_{\pi}+m_{N})}$} where $s_R=(m_\pi+m_N)^2$ being the physical $\pi N$ threshold. Naturally, one expects $R$ being the interaction range~\cite{Regge:1958ft}, $i.e.$, $R\simeq 0.52m_\pi^{-1}\simeq 0.79$ fm for $\pi N$ scatterings~\cite{Wang:2018nwi}. Notice that with Eq.~(\ref{f}) we can actually define different contribution from different energy region to the background term, through dividing the integral interval. For example, by calculating $f^{\chi PT}(s)\propto \int_{-\Lambda^2_{\chi PT}}^{s_L}$, we may estimate the contribution from the energy region calculable using perturbation theory~\cite{Alarcon:2012kn,Chen:2012nx,Siemens:2016hdi,Siemens:2016jwj,Siemens:2017opr}: if $\Lambda^2_{\chi PT}$ is set as $0.08$ GeV$^2$ according to the $N^*(1440)$ pole location~\cite{Wang:2018nwi}, then $R_{\chi PT}=0.39$ fm\footnote{As already discussed in Ref.~\cite{Wang:2018nwi}, $\Lambda^2_{\chi PT}=0.08$ GeV$^2$ is a bit too optimistic, however the fit requires a larger $R$ value. }.

It is actually easy to understand why $R$ becomes too small in Fit I. Since the only source of branch point singularity in $S_{11}$ before the inelastic effect in $\rho_2$ is switched on, is $i\rho_1(s)$, and remember that on the left cut $i\rho_1(s+i\epsilon)=-i\rho_1(s-i\epsilon)$, the $S$ matrix constructed as such is also unitary on the left! Hence the spectral function of the left hand integral vanishes. This annoying property of $S$ is due to the over-simplification of Eqs.~(\ref{case1-kmatrix}), (\ref{fit1}): according to Cutkosky rule, the function $i\rho(s)$ comes from the absorptive part of two body intermediate states, thus it should have come along with its dispersive or real part missed in Eq.~(\ref{fit1}), which is
widely used in the literature. Therefore to cure the deficiency one substitutes $i\rho(s)$ in those equations by the well known two point function (also known as Chew-Mandelstam function) $B_0(s;m_1^2,m_2^2)$ defined as following ($b_0$ is a subtraction constant),
\begin{equation}\label{B0def}
B_0(s;m_1^2,m_2^2)={b_0+\frac{s}{\pi}}\int ds'\frac{\rho(s')}{s'(s'-s)}\ .
\end{equation}

\begin{flushleft}
\textbf{Fit II:}
\end{flushleft}

As discussed above, the vanishing of the left hand cut contribution comes from a poor approximation (on-shell approximation). In order to parameterize the amplitude with better analyticity property, one should at least pick up the dispersive part together with the absorptive part.
{From Eq.~\eqref{B0def}}, it reads as, up to a {subtraction} constant,
\begin{equation}\label{B0}
  \begin{aligned}
  B_{0}(s,m_{1}^2,m_{2}^2)=&\frac{1}{\pi}
  +\frac{1}{2\pi s}\{[s(1+\rho(s))-\Delta]\mathrm{ln}[\frac{\Delta+s(1-\rho(s))}{\Delta-s(1+\rho(s))} ]\\
  &+[s(1-\rho(s))-\Delta]\mathrm{ln}[\frac{\Delta+s(1+\rho(s))}{\Delta-s(1-\rho(s))} ]	\}\ ,
  \end{aligned}
\end{equation}
where {$\Delta=m_N^2 - m_{\pi}^2$} for example.
This function has the same imaginary part as $i\rho(s)$ in the physical region, but maintains quite different analyticity property: it is analytic on the left while $i\rho(s)$ contains a left cut starting from {$s_L=(m_N-m_\pi)^2$} to $-\infty$.
	
We therefore unitarize the amplitude in the following way:
\begin{equation}\label{t-matrix-b0}
{T^{-1} = K^{-1} -B\ ,\quad B\equiv\text{diag}\{B_0(s,m_\pi^2,m_N^2),\ B_0(s,m_\eta^2,m_N^2)\}}\ ,
\end{equation}
and parameterize the $K$-matrix as before, and fit data of $\pi N$ scattering amplitude from threshold to $2.1\mathrm{GeV}$. The fit is also plotted in Fig.~\ref{figcase1}.

\begin{figure}[h]
	\centering
	\includegraphics[width=0.65\textwidth]{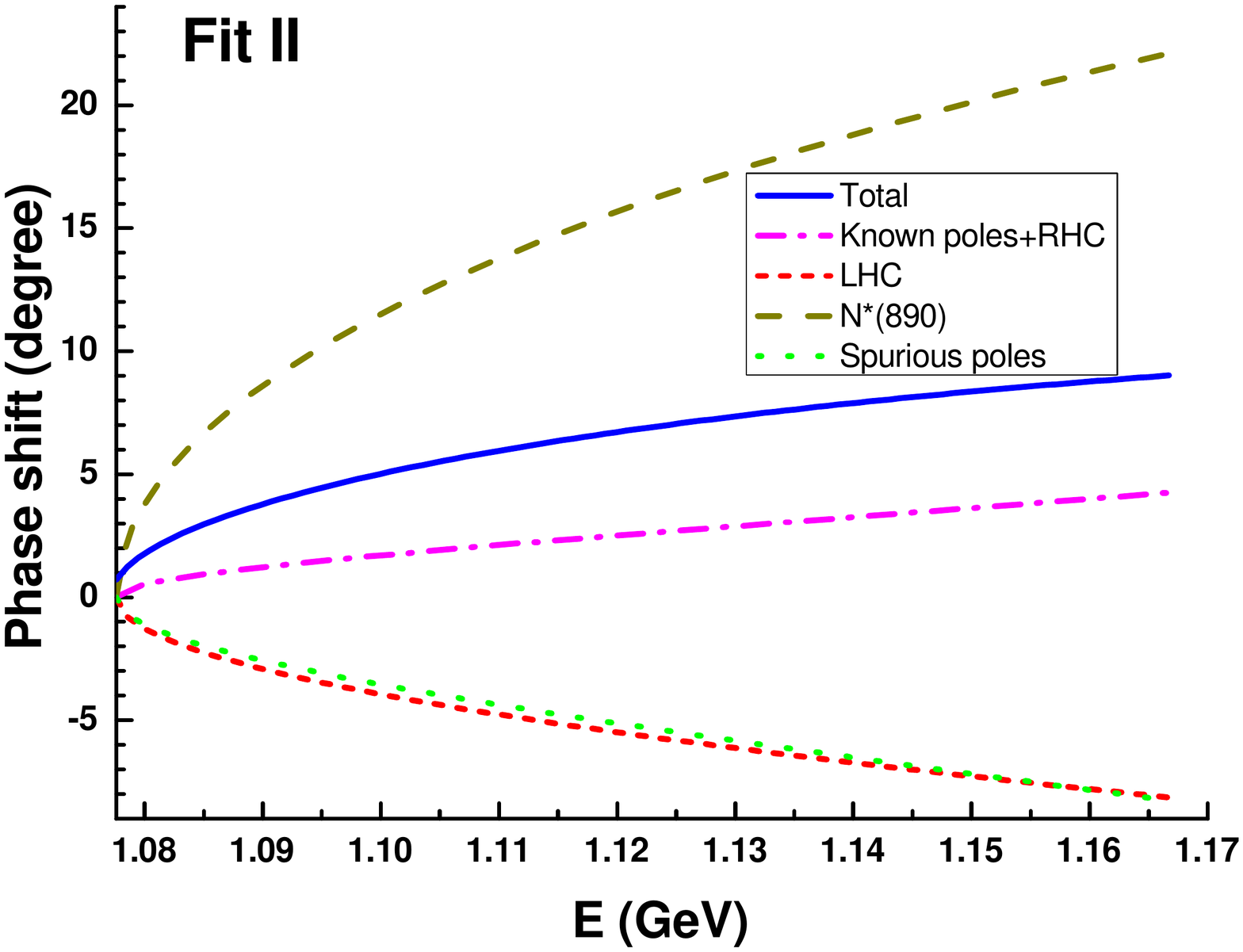}
	\caption{Different contributions to the phase shift in Fit II.}
	\label{phase-2}
\end{figure}

Nearby poles found from Fit II are listed in Table~\ref{tabcase2}. We also plot {$\mathrm{Im}(\frac{\ln S_{11}(s')}{2i\rho_1(s')})$} read off from Fit II in Fig.~\ref{LHC-1}. The scattering length does not change much: $a\simeq 164.11\times 10^{-3}m_\pi^{-1}$.
We find  significant improvement on the left-hand cut. It has now a similar behavior with the $\chi PT$ calculation in the validity region of the latter, meanwhile contributes still negatively to the integral of $f$ defined in Eq.~(\ref{f}):
in the region from 0 to $s_L = (m_N-m_{\pi})^2$, the function $\mathrm{Im}(\frac{\ln S_{11}(s')}{2i\rho_1(s')})$ is positive definite and $\frac{1}{s'(s'-s)}$ is negative, hence according to Eq.~(\ref{f}), the contribution to the integral in this region is negative; while in the region from $-\infty$ to 0, $\mathrm{Im}(\frac{\ln S_{11}(s')}{2i\rho_1(s')})$ is negative and $\frac{1}{s'(s'-s)}$ is positive. One gets $R=0.18$ fm in Fit II which is not as large as what is required by fit~\cite{Wang:2018nwi}, but is much improved comparing with the $R (=0)$ value in Fit I.
Also, there is an overall improvement of fit quality comparing with that of Fit I, in the sense that the spurious pole is pushed further away as its contribution to the phase shift gets much smaller, as can be seen by comparing Figures~\ref{phase-1} and \ref{phase-2}.
\begin{table}[h]
    \caption{Fit II: Poles obtained using Eq.~(\ref{t-matrix-b0})}
    \centering
    \begin{tabular}{p{2cm}p{8cm}}
        \hline
         Sheets & pole positions on $\sqrt{s}$ plane in unit of GeV \\
        \hline
		$\mathrm{I}$              &$1.85 - 0.29i$\\
	  \hline	
$\mathrm{II}$                 & $1.66- 0.04i$; $0.77-  0.23i$ \\
	  \hline	$\mathrm{III}$              & $1.66- 0.06i$;  $1.50-  0.05i$ 
 \\
  \hline		$\mathrm{IV}$              & $1.53-0.01i$\\
		\hline
    \end{tabular}
    \label{tabcase2}
\end{table}

Comparing Fit II with  Fit I we observe that the correct use of analyticity in the $s$ channel, by recovering the dispersive part of the kinematic factor, leads to much improved predictions on the left cuts, the much suppressed spurious pole contribution, and stronger evidence for the existence of $N^*(890)$. Inspired by this, one wonder whether the inclusion of $t$--channel and $u$--channel cuts could further improve the fit results. We do this in the following Fit III.

\begin{flushleft}
	\textbf{Fit III:}
\end{flushleft}

\begin{figure}[h]
	\centering
	\includegraphics[width=0.65\textwidth]{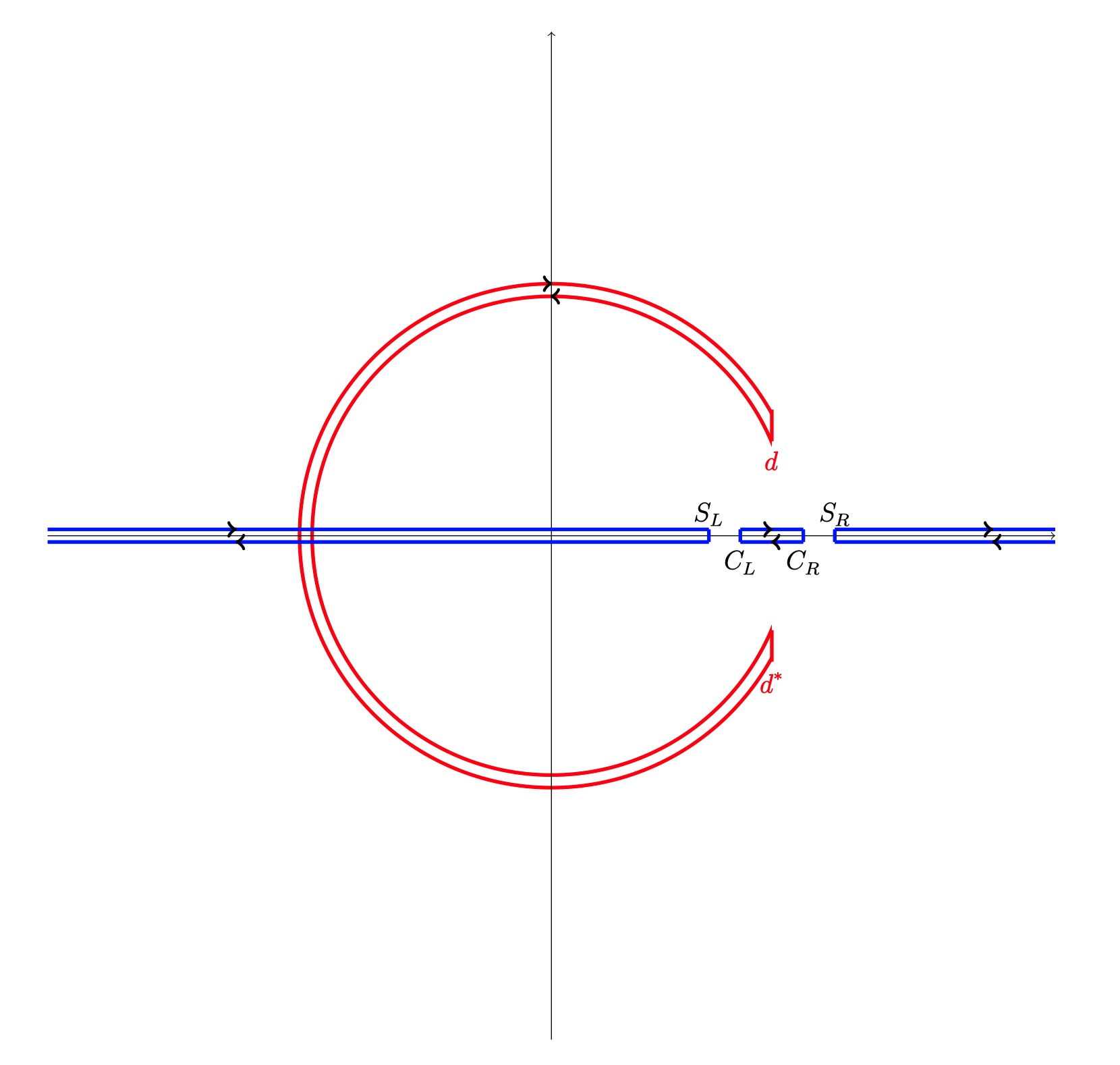}
	\caption{The left cut caused by $t$--channel $\rho$ meson exchange (circular arc); $u$-channel exchange (line segment from $c_L$ to $c_R$ \cite{Wang:2018nwi}).}
	\label{rhocut}
\end{figure}

We add in the $K$-matrix the tree diagram contribution of $t$-channel $\rho$ meson and $u$-channel nucleon exchanges (though the latter's effect is very small).
It now reads,
\begin{equation}\label{fitiii}
  K_{ij}=\sum_{\alpha=1,2}\frac{c^{\alpha}_{ij}}{s-M^2_{\alpha}}+P^{(2)}_{ij}(s)+K^{t,u-channel }_{ij}\ ,
\end{equation}
and T-matrix parameterized as Eq.~(\ref{t-matrix-b0}). Now the ``left" cut becomes rather complicated, see Fig.~\ref{rhocut}.
We leave the explanation in the appendix. Fit results are again plotted in Fig.~\ref{figcase1}.
Nearby poles on four sheets are shown in Table~\ref{tabcase3}. The scattering length still does not change much: $a\simeq 167.08\times 10^{-3}m_\pi^{-1}$.
\vspace{0cm}
\begin{table}[h]
    \caption{Fit III: Poles obtained using Eq.~(\ref{fitiii})}
    \centering
    \begin{tabular}{p{2cm}p{8cm}}
        \hline
         Sheets & pole positions on $\sqrt{s}$ plane in unit of GeV.) \\
        \hline
		$\mathrm{I}$              & $1.35-0.58i$\\
	  \hline	
        $\mathrm{II}$                 &  $1.67- 0.07i$; $0.93-  0.27i$   \\
	  \hline	$\mathrm{III}$              & $1.65-0.09i$; $1.53- 0.07i$     \\
  \hline		$\mathrm{IV}$              & $1.54-0.01i$  \\
		\hline
    \end{tabular}
    \label{tabcase3}
\end{table}
The spectral function  read off from Fit III is plotted in Fig.~\ref{LHC-1}.
It is seen that the value of the spectral function now gets more closed to the result in Ref.~\cite{Wang:2018nwi}, comparing with the Fit II solution in the small $|s|$ region, in the sense that it also produces the circular cut. It is even more important to stress that, the investigation here further justifies the strategy adopted in Refs.~\cite{Wang:2017agd,Wang:2018gul,Wang:2018nwi}, i.e., to use a cutoff parameter to regulate the ``resonance region" contribution to $f(s)$. Since the ``resonance region" contribution gives the same sign as comparing with that from perturbation region, the strategy taken in Refs.~\cite{Wang:2017agd,Wang:2018gul,Wang:2018nwi} won't be  bad, as the fit decides the cutoff parameter and, after all,  the location of the $N^*(890)$ pole won't be annoyed much by such an uncertainty. Numerically, here we obtain
$R(\mathrm{left~cut})\simeq 0.25 fm$, $R(\mathrm{circular~arc})\simeq 0.10 fm$, their sum is closer to the fit value obtained in Ref.~\cite{Wang:2018nwi}.
 Also,
the resonance pole below threshold  provides a larger phase shift. Nevertheless
  the contribution of spurious pole is not found to be further  suppressed comparing with Fit II.
\begin{figure}[h]
	\centering
	\includegraphics[width=0.65\textwidth]{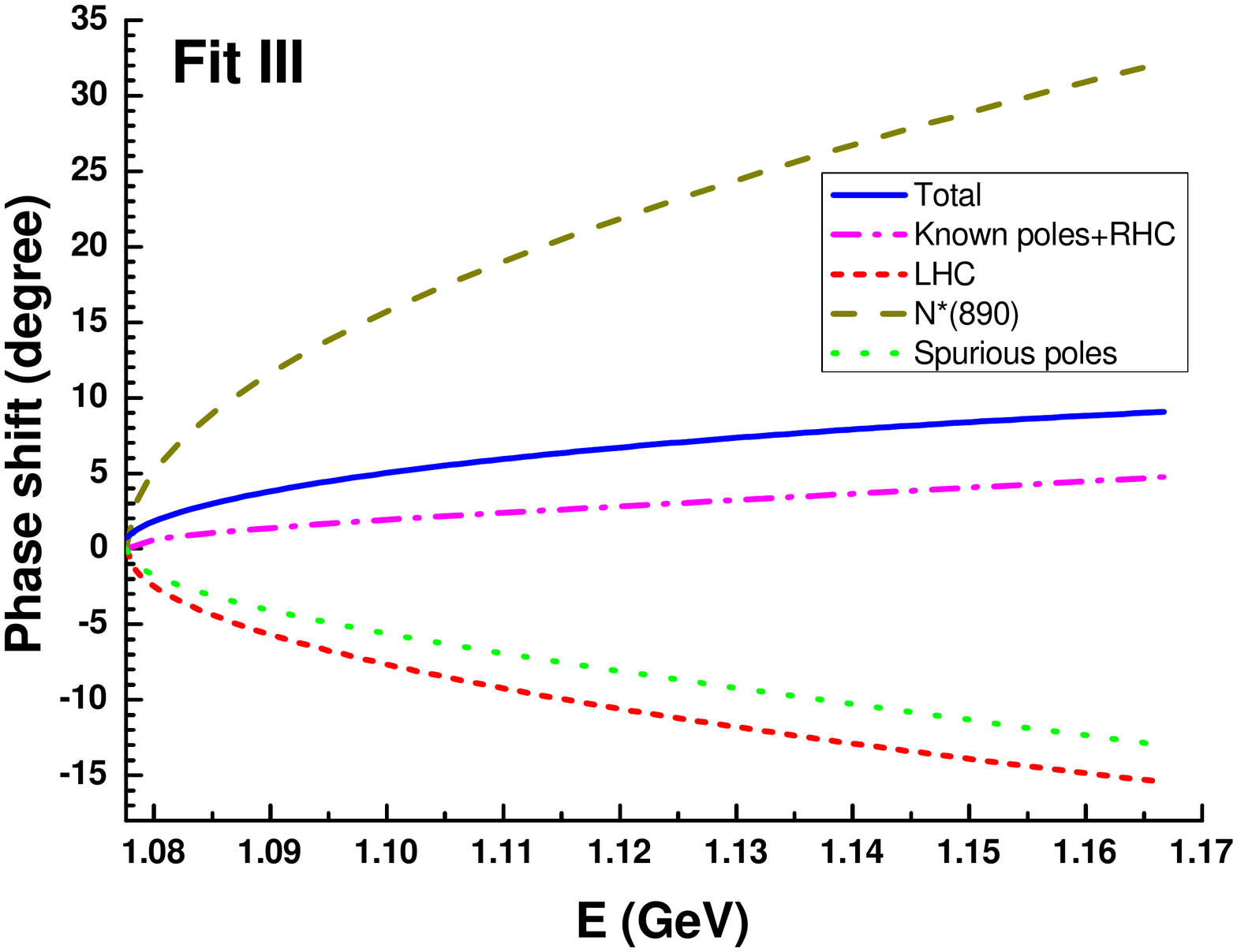}
	\caption{Fit III: The phase shift near $\pi N$ threshold.}
	\label{phase-3}
\end{figure}

At last one would like to fit the data by getting rid of spurious poles and by borrowing the left cuts (both the cut $(-\infty, s_L]$ and the circular arc), i.e., to use predictions on ``left hand cuts" as an input in applying the PKU representation. The fit to the data is achieved without the need of any spurious poles and cut-off parameters. In this way one gets the subthreshold pole located at $\sqrt{s}\simeq 796(2)-109(8)i$MeV -- a result compatible with the results of Refs.~\cite{Wang:2018gul,Wang:2018nwi}, even though the $R$ value used here is only half of the value as estimated in Ref.~\cite{Wang:2018nwi}. Note that the location of $N^*(890)$ pole here numerically may not be necessarily better than the result in Ref.~\cite{Wang:2018nwi}; actually the aim of this paper is to show in another way that the strategy of Ref.~\cite{Wang:2018nwi} and the existence of $N^*(890)$, are reasonable.

To summarize, using a simple unitarization model,
in this paper we have shown that, a better treatment of analyticity in the $s$ channel dynamics has led to much improved analytic property on the left side. The inclusion of crossing symmetry, i.e., $t$ and $u$ channel resonance exchanges has further improved the quality of its predictions, in the sense that the results get closer to the $\chi PT$ ones in the validity region of the latter, as well as the emergence of the circular cut. Using the ``best" model predictions on the left cuts (i.e., Fit III), one gets the pole location of the $N^*(890)$ being consistent with that of Refs.~\cite{Wang:2017agd,Wang:2018gul,Wang:2018nwi}. And the advantage now is the elimination of any cutoff dependence when evaluating left cut integrals. We think it is important to stress that all the pseudo-thresholds, i.e. $(m_1-m_2)^2$
are essentially due to  $relativistic$ effects and should not be ignored. In a non-relativistic theory, one may take these effects into account through introducing a sizable, physical, and negative background contribution to the phase shift~\cite{Regge:1958ft}.

We notice that in a paper by D\"{o}ring and Nakayama~\cite{Doring:2009uc} (see also Ref.~\cite{Doring:2009yv}), the authors found a sub-threshold pole at $\sqrt{s}\simeq 1031-203i$ MeV in $S_{11}$ channel, without further information on left cuts and spurious poles. The authors carefully quote:
\textit{``However, it is not clear if this state is genuine or a forced pole that mocks up the u- and t-channel subthreshold cuts that are not explicitly included in the present model. ''}The investigations of this paper and Refs.~\cite{Wang:2017agd,Wang:2018gul,Wang:2018nwi} made it clear, we think, that the subthreshold resonance does not play a role of mocking up the crossed channels effects. The fact is just on the contrary -- the former contributes a positive definite phase shift to counter balance the effects of the latter.
We have actually {witnessed} similar things that happened in $\pi\pi$ scatterings and $\pi K$ scatterings, in most attractive channels~\cite{Xiao:2000kx, Zheng:2003rw,Zhou:2004ms,Zhou:2006wm}.

\vspace{1cm}
\textit{Acknowledgement:} This work is supported in part by National Nature Science Foundations of China under contract number 11975028
and  10925522 
, and by the Sino-German CRC 110 (Grant No. TRR110). The authors would like to thank Ulf-G.~Mei{\ss}ner for helpful discussions and suggestions.

\renewcommand\refname{Reference}
\bibliographystyle{h-physrev}
\bibliography{niuye}

\newpage
\begin{center}
\Large{Appendix}
\end{center}

To calculate the background contribution $f(s)$ using dispersion relations, we need to determine the analytical structure of $f(s)$. In Fit I and Fit II, the left hand cut only lies on real axis. In Fit III, the $\rho$ meson exchange in $t$-channel introduces a circular arc cut in $s$-plane, as depicted in Fig.~\ref{rhocut}, which actually coincides with the cut generated by $t$ channel $\pi\pi$ continuous spectrum~\cite{Kennedy1962}. Here we give a detailed explanation on this.

The relevant Lagrangians can be found, for example in Ref.~\cite{Lahiff:1999ur}. The $S_{11}$ $\rho$ exchange amplitude is,
\begin{equation}
    \begin{aligned}
      T(S_{11}) =& 2(T^{-,1/2}_{++} + T^{-,1/2}_{+-})\ ,\\
      T^{-,1/2}_{++}=&(s-m^{2}_{\pi}-m^2_{N})B^{-,1/2}_{C}\ ,\\
      T^{-,1/2}_{+-}=&-\frac{m_N}{\sqrt{s}} (s+m^{2}_{\pi}-m^2_{N})B^{-,1/2}_{S}\ ,\\
      B^{-,1/2}_{C}=&-\frac{g_{\rho NN}g_{\rho \pi\pi}}{64\pi} I^{t}_{C}(m_3)\ ,\\
	  B^{-,1/2}_{S}=&-\frac{g_{\rho NN}g_{\rho \pi\pi}}{64\pi} I^{t}_{S}(m_3)\ ,
    \end{aligned}
\end{equation}
where $m_3$ is the mass of the exchanged particle.
To determine the location of circular cut we need to look into the expressions of $I^{t}_{C}(m_3)$ and $I^{t}_{S}(m_3)$:
\begin{equation}\label{Ictdef}
\begin{aligned}
      I^{t}_{C}(m_{3}) =&\int^{1}_{-1}\frac{1+z_s}{t-m_{3}^{2}}dz_{s}
      =\frac{4}{\rho^{4}s^2}\left[\rho^2s + T(m_3,s)\ln{\frac{m_{3}^{2}}{T(m_3,s)} }\right]\ ,\\
      T(m_3,s)=& \frac{s^2-2(m_N^2+m_{\pi}^2)s+m_3^2s+(m_N^2-m_{\pi}^2)^2}{s}\ ,
\end{aligned}
\end{equation}
and $I^{t}_{S}$ behaves similarly. The logarithmic function offers a discontinuity when the phase of its argument equals to $\pi$, i.e.
\begin{equation}\label{Tnega}
      T(m_3,s)= \frac{s^2-2(m_N^2+m_{\pi}^2)s+m_3^2s+(m_N^2-m_{\pi}^2)^2}{s} \in \mathbb{R}^{-}\ .
\end{equation}
The numerator of $T(m_3,s)$ is a quadric expression which has two complex roots when $2m_{\pi}<m_3<2m_{N}$:
\begin{equation}\label{ddsdef}
\begin{aligned}
      d =& m^{2}_{N}+m_{\pi}^2-\frac{m_{\rho}^2}{2}
       + i\sqrt{m_{\rho}^{2}(m_{N}^2+m_{\pi}^2)-4m_{N}^2m_{\pi}^2-\frac{m_{\rho}^4}{4}}\ ,\\
      d^{*} =& m^{2}_{N}+m_{\pi}^2-\frac{m_{\rho}^2}{2}
       - i\sqrt{m_{\rho}^{2}(m_{N}^2+m_{\pi}^2)-4m_{N}^2m_{\pi}^2-\frac{m_{\rho}^4}{4}}\ ,\\
      |d| = & (m_{N}^2 - m_{\pi}^2)\ ,
\end{aligned}
\end{equation}
so the logarithmic term in Eq.~\eqref{Ictdef} can be rewritten as $\ln[(s-d)(s-d^{*})/s]$. Defining
\begin{equation}
    \begin{aligned}
      d =& a +ib\ ,\\
      s =& x + iy\ ,
    \end{aligned}
\end{equation}
the argument becomes
\begin{equation}
	\frac{(s-d)(s-d^{*})}{s}
	=\frac{(x-a)^2-y^2+b^2+2iy(x-a)}{x+iy}\ .
\end{equation}
The requirement of Eq.~\eqref{Tnega} yields
\begin{eqnarray}
      && x^2 + y^2 = a^2 + b^2\ \,\,\,\mathrm{and}\,\,\,\ x<a\ ,\nonumber\\
      && \mathrm{or}\,\,\,\ y = 0\ \,\,\,\mathrm{and}\,\,\,\ x<0\ ,\label{cutsrho}
\end{eqnarray}
which implies that in the cut off real-axis,
\begin{equation}
    |s| = m_N^2 - m_{\pi}^2\ ,
\end{equation}
namely the complex cut led by $t$-channel $\rho$ exchange should lie on the circular arc centered at the origin with a radius $m_N^2 - m_{\pi}^2$. As shown in Fig.~\ref{rhocut},  The $S$-matrix has a discontinuity along the circular arc between $d$ and $d^{*}$ in Eq.~\eqref{ddsdef}.

Eq.~\eqref{cutsrho} shows another cut lying on the real axis from $-\infty$ to $0$, which is covered by l.h.c from $-\infty$ to pseudo threshold. As pointed out in Ref.~\cite{Kennedy1962}, this always appears as a trivial solution of the left-hand singularities.
\end{document}